\pdfoutput=1
\documentclass[conference]{IEEEtran}

\IEEEoverridecommandlockouts

\usepackage{graphicx}
\usepackage{amsmath}
\usepackage{siunitx}
\usepackage{cite}
\usepackage{stfloats}
\usepackage{hyperref}
\usepackage{cuted}

\usepackage{etoolbox}
\usepackage{eso-pic}

\makeatletter
\patchcmd{\@maketitle}{\vskip 1.0em}{\vskip 0.2em}{}{}
\patchcmd{\@maketitle}{\vskip 1.0em}{\vskip 0.2em}{}{}
\makeatother

\begin{document}

% --- arXiv preprint: IEEE copyright notice on first page only ---
\AddToShipoutPictureFG*{%
  \AtPageLowerLeft{\put(36,18){%
    \parbox{\dimexpr\paperwidth-72pt\relax}{%
      \scriptsize\copyright~2026 IEEE. Personal use of this material is permitted. Permission from IEEE must be obtained for all other uses, including reprinting/republishing this material for advertising or promotional purposes, creating new collective works, for resale or redistribution to servers or lists, or reuse of any copyrighted component of this work in other works.}%
  }}%
}
% ---------------------------------------------------------------
\bstctlcite{IEEEexample:BSTcontrol}

\title{Multi-Channel Soil Moisture Measurement: High Accuracy and Low Crosstalk Through Optical-Semiconductor Based Differential Sensing}

\author{
\IEEEauthorblockN{Thomas Maier, Charlotte Rohleder}
\IEEEauthorblockA{
\textit{Machine Learning and Data Analytics Lab (MaD-Lab)}\\
\textit{Friedrich-Alexander-Universität Erlangen-Nürnberg (FAU)}\\
\{thomas.maier, charlotte.pradel\}@fau.de}
\and
\IEEEauthorblockN{Lukas Kamm, Philipp Dauner}
\IEEEauthorblockA{
\textit{Agvolution GmbH}\\
\textit{Erlangen, Germany}\\
\{l.kamm, p.dauner\}@agvolution.com}
\and
\IEEEauthorblockN{Jan Pinski}
\IEEEauthorblockA{
\textit{Department of Environment and}\\
\textit{Urban Greenery}\\
\textit{State Capital Hanover}\\
www.hannover.de/umwelt-stadtgruen\\
jan.pinski@hannover-stadt.de}
\and
\IEEEauthorblockN{Bjoern M. Eskofier}
\IEEEauthorblockA{
\textit{Chair of AI-supported Therapy Decisions, LMU München, Munich, Germany}\\
\textit{Department Artificial Intelligence in Biomedical Engineering (AIBE)}, Germany\\
\textit{Friedrich-Alexander-Universität Erlangen-Nürnberg (FAU), Erlangen, Germany}\\
\textit{Institute of AI for Health, Helmholtz Zentrum München, Neuherberg, Germany}\\
bjoern.eskofier@fau.de}
}

\maketitle

\begin{abstract}
Soil moisture measurement plays a key role in irrigation and environmental management. Yet it remains unreliable due to heterogeneous soils, limited sensing volumes, temperature drift, and parasitic inter-channel coupling. This work presents a compact multi-depth capacitive probe that extends a parallel-plate geometry from previous work with differential activation to suppress stray capacitances and improve accuracy. An equivalent-circuit model quantifies parasitic effects, and optically coupled transistor bridges isolate each sensing layer. Raw capacitance is converted to volumetric water content and plant-available water using established calibration models. Laboratory results show a fourfold reduction in temperature sensitivity, strong confinement of the sensing volume, and improved repeatability in heterogeneous soils. Field validation against reference sensors demonstrates high accuracy and precision comparable to widely used instruments, enabling a practical and scalable solution for agricultural and urban soil-moisture monitoring.
\end{abstract}


\section{Introduction}
\begin{figure*}[!b]
  \centering
  \vspace{-12pt}
  \includegraphics[width=\textwidth]{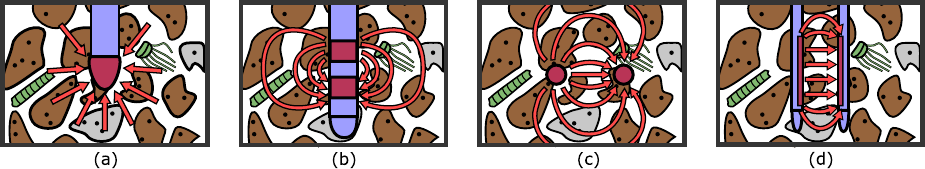}
  \vspace{-20pt}
  \caption{Typical sensor geometries and sampling volumes in heterogeneous soil: 
  (a) tensiometer \cite{Richards1942}, 
  (b) ring-electrode FDR probe \cite{Babaeian2019}, 
  (c) TDR wires \cite{Annan1977}, 
  (d) parallel-plate probe enabling homogeneous sampling \cite{Maier2023}.}
  \label{fig:homogSensing}
\end{figure*}
\vspace{-3pt}

Climate change alters global precipitation and demands more efficient water use in agriculture and urban greenery \cite{Trenberth2011}. Accurate soil-moisture monitoring enables optimized irrigation but remains challenging due to the need for site-specific calibration and poor reproducibility \cite{Jackisch2020, Leib2003, Walker2004, Rowlandson2013}. Reported limitations include soil disturbance during installation \cite{Evett2002, Teixeira2003}, dependence on texture and bulk density \cite{Gong2003, Baumhardt2000}, small or uneven sensing volumes \cite{Ranjan1997, Nissen2003, Annan1977, Maier2017}, and temperature-related drift \cite{Saito2009, Pepin1995, Decagon2009, Nieberding2023}. As a result, commercial probes often fail to provide consistent readings in heterogeneous soils, and satellite missions remain limited to surface layers \cite{Gorrab2014, Ding2023}.
In practice, in-situ sensor networks combined with IoT technologies are used to capture vertical profiles \cite{Ullo2020}. These rely on electrical, dielectric, thermal, or tensiometric principles \cite{Kawahara2012, Song1998, Noborio1996, Owen1961, Loewer2017, Topp1980}, where raw signals (voltage, frequency, counts) are converted to volumetric water content (VWC) or soil-water potential \cite{Kanoun2004, Topp1980}. VWC is the water–volume ratio \cite{Gardner1998, Topp1980}, while soil-water potential quantifies suction forces of the soil \cite{Cassel1986}. For irrigation control, values are often normalized as plant-available water (PAW), bounded by field capacity and wilting point \cite{Cassel1986, Datta2017, Kramer1995}. Because these thresholds vary strongly with texture, from about \SI{5}{\percent} VWC for sand to $>$ \SI{30}{\percent} for clay \cite{Kramer1995}, soil-specific calibration remains necessary \cite{Topp1980}. The absence of a universal conversion standard is still a major obstacle to comparable soil-moisture data.
Soil-moisture information supports precision irrigation and ecosystem resilience in both agriculture and cities \cite{Kropp2019, Stratopoulos2019, Schuett2022a}. Because over \SI{75}{\percent} of the roots of annual crops and young city trees occur within \SI{60}{\centi\meter} of the surface \cite{Jackson1996, Fan2016}, sensors covering this range capture most root-zone dynamics while avoiding compact subsoil \cite{Armstrong2009d}. Natural soils are stratified in texture and density \cite{Cassel1986, Armstrong2009a}, requiring sensing volumes that balance spatial noise and layer separation \cite{Baveye2002,Whitaker1969, Wu2020, CostanzaRobinson2011}.The probe geometry in this work is chosen to balance these effects and provide stable, layer-specific measurements.
Conventional probes exhibit non-uniform sensing fields biased toward the sensor surface. Tensiometers measure only at the ceramic tip, see Fig. \ref{fig:homogSensing} (a) \cite{Richards1942}. Rod-type capacitive and TDR sensors concentrate fringing fields near electrodes, see Fig. \ref{fig:homogSensing} (b),(c) \cite{Kojima2016, Shirahama2015, Knight1997, Zegelin1989}. Water films or compaction during insertion may distort readings \cite{Nissen2003, Limsuwat2009, Bell1987, Gardner1998, Teixeira2003}. To achieve homogeneous sampling, our previous design \cite{Maier2023} used two parallel-plate electrodes inserted from the surface, generating a uniform field through a defined soil slab and allowing multi-depth measurements, see Fig. \ref{fig:homogSensing} (d) \cite{Maier2023}. This paper presents an improved soil-moisture sensor addressing these issues. A differential sensing scheme sequentially isolates electrodes using optically coupled MOSFET bridges, subtracting stray and coupling capacitances verified by a circuit model and node-voltage analysis. In addition, a conversion from corrected capacitance to normalized PAW using soil-specific field-capacity and wilting-point parameters together with laboratory and field validation is presented.

\section{Materials and Methods}
\label{sec:MatAndMet}
\subsection{Equivalent circuit model}
\label{subsec:ECM}

\begin{figure}[!t]
  \centering
  \includegraphics[width=0.35\textwidth]{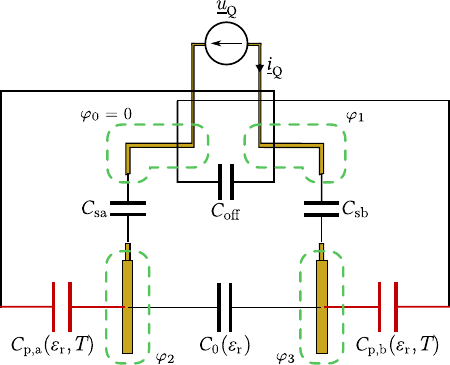}
  \caption{Equivalent circuit diagram used for calculations. 
  $C_\text{sa}$ and $C_\text{sb}$ represent the electronic switches as capacitors, while $C_\text{p,a/b}(\varepsilon_\text{r}, T)$ denote the parasitic coupling capacitors. Dashed lines indicate the nodes used for nodal analysis, where $\varphi_2$ \& $\varphi_3$ are the measuring electrodes of a channel and $\varphi_0$ \& $\varphi_1$ are the connecting traces between the excitation signal $\underline{u}_\text{Q}$ and electronic switches.}
  \vspace{-10pt}
  \label{fig:ESBmodelling}
\end{figure}

Embedding several capacitive sensing zones for multi-depth measurements introduces stray coupling and wiring parasitics between adjacent electrodes, causing inter-channel crosstalk \cite{Clayton1982}, which is quantified using the equivalent-circuit model in Fig.~\ref{fig:ESBmodelling}.
Each channel measures the electrode–soil capacitance $C_{\text{0}}(\varepsilon_\text{r})$, formed by the insulation capacitances $C_{\text{iso}}$ on both electrodes in series with the soil capacitance $C_{\text{r}}(\varepsilon_\text{r})$, plus a fixed wiring offset $C_{\text{off}}$. 
Parasitic couplings to neighbouring channels and the ground shield are modelled as $C_{p,a}(\varepsilon_\text{r}, T)$ and $C_{p,b}(\varepsilon_\text{r}, T)$, which depend on relative soil permittivity $\varepsilon_\text{r}$ and temperature $T$. 
To enable subsequent crosstalk compensation, ideal switches, modelled as capacitors $C_{sa}$ \& $C_{sb}$, are included and are first assumed closed (infinite capacitance value). 
The excitation signal $\underline{u}_{\mathrm{Q}}$ is generated by a capacitance-to-digital converter, which drives the sensor with a sinusoidal signal of $1.2\,\mathrm{V}_{\mathrm{pp}}$ at \SIrange{1}{3}{\mega\hertz} using an internal LC resonant network and measures resonant frequency shifts to infer capacitance changes. 
The digitized values are read by an on-board microcontroller for capacitance conversion.

\subsection{Mapping measured capacity to normalized soil permittivity}
\label{subsec:TranslationBetween}
$C_{\text{0}}(\varepsilon_\text{r})$ can be expressed as a series connection of the electrode insulation capacitances and the soil capacitance \cite{DiLorenzo2020, Maier2017, Kelleners2004, Escriba2020, Gardner1998}:
\begin{equation}
\label{eq:seriesCap}
    C_{\text{0}}(\varepsilon_\text{r}) = \bigg(\frac{2}{C_{\text{iso}}} + \frac{1}{\varepsilon_{\text{r}} C_{\text{r}}} \bigg)^{-1} + C_{\text{off}} .
\end{equation}

Higher $C_{\text{iso}}$ (thinner insulation) or larger $C_{\text{r}}(\varepsilon_\text{r})$ values increase the sensor’s dynamic range but also its nonlinearity. 
For very thin insulations  below \SI{30}{\micro\meter}, insulation permittivity varies due to moisture absorption \cite{Maier2017}. 
This effect is negligible for the \SI{1}{\milli\meter}–\SI{1.5}{\milli\meter} layers used in this setup.
The capacitance is normalized using min–max scaling to obtain $C_{\text{0}}(\varepsilon_\text{r})'$:
\begin{equation}
    C_{\text{0}}(\varepsilon_\text{r})' = 
    \frac{C_{\text{0}}(\varepsilon_\text{r}) - C_{\text{0}}(\varepsilon_\text{r}(\text{air}))}
         {C_{\text{0}}(\varepsilon_\text{r}(\text{water})) - C_{\text{0}}(\varepsilon_\text{r}(\text{air}))},
\end{equation}
where $\varepsilon_\text{r}(\text{water})=81$ and $\varepsilon_\text{r}(\text{air})=1$.
This normalization ensures comparability across probes and materials\cite{Bogena2017, Robinson2003}.  
To generalize calibration, $\varepsilon_\text{r}$ is also normalized as ${\varepsilon_\text{r}}'$ and related to $C_{\text{0}}({\varepsilon_\text{r}}')'$ through:
\begin{equation}
\label{eq:generalForm}
     C_{\text{0}}({\varepsilon_\text{r}}')' = \bigg(a+\frac{b}{{\varepsilon_\text{r}}'} \bigg)^{-1} +c ,
\end{equation}
with $a=2/C_{\text{iso}}$, $b=1/C_{\text{r}}(\varepsilon_\text{r})$, and calibration conditions $C_{\text{0}}(0)'=0$, $C_{\text{0}}(1)'=1$, giving $c=0$, $b=1-a$. 
Solving for  ${\varepsilon_\text{r}}'$ yields:
\begin{equation}
\label{eq:normPerm}
    {\varepsilon_\text{r}}' = \frac{C_{\text{0}}({\varepsilon_\text{r}}')'(a-1)}{a\,C_{\text{0}}({\varepsilon_\text{r}}')'-1}.
\end{equation}
Thus, an additional measurement in ethanol ($\varepsilon_\text{r}=23$) completes the three point calibration (air-ethanol-water), determines $a$, compensates manufacturing tolerances, and, after over 200 calibrations, has proven statistically stable, eliminating the need for per sensor calibration.
For embedded implementation, Eq.~(\ref{eq:normPerm}) was approximated by a piecewise linear model with $<10^{-3}$ error, reducing computational load. 

\subsection{Translation to VWC and PAW}
\label{subsec:PAWTranslation}
After calibrating the normalized permittivity ${\varepsilon_\text{r}}'$, the next step is to derive the VWC-PAW relation. 
Although the Topp equation \cite{Topp1980} is widely used, it is based on time-domain reflectometry at higher frequencies and is not directly applicable to the capacitive setup here. 
Several studies investigated the dependence of soil permittivity on water content while considering bulk density, texture, organic matter, and porosity \cite{Gardner1998, Roth1992}. 
Among these, the Gardner model \cite{Gardner1998} best matches both the operating frequency range and the need for parameterization across different soils. 
The VWC is calculated as
\begin{equation}
\label{eq:gardnerVWCModel}
\theta = \frac{\sqrt{\varepsilon_\text{r}} + 1.208 - 2.454 \cdot \rho}{9.93},
\end{equation}
where $\rho$ is the bulk density, determined individually for each soil layer by drill-stick sampling. 
Accordingly, the VWC–PAW conversion is soil-specific and requires layer-wise bulk-density measurements for each installation site. 
Effects of salinity and organic matter are primarily mediated through changes in effective permittivity and are therefore captured indirectly in the dielectric response. 
The available water capacity (AWC) is defined as the difference between the VWC at field capacity~$\theta_\mathrm{fc}$ and at the permanent wilting point~$\theta_\mathrm{pwp}$. 
Normalizing $\theta$ to this range yields the PAW value:
\begin{equation}
\label{eq:plantAvailableWater}
\mathrm{PAW} = \frac{\theta}{\theta_\mathrm{awc}} = 
\frac{\theta}{\theta_\mathrm{fc} - \theta_\mathrm{pwp}} .
\end{equation}
For technical or non-natural substrates, laboratory soil analyses can provide these parameters from undisturbed samples, but require costly sampling, processing, and site-specific database integration \cite{Maier2025}. Therefore, literature-based estimates for $\theta_\mathrm{fc}$ and $\theta_\mathrm{pwp}$ are taken from \cite{Wesselok2014} as first approximations and subsequently normalized using the seasonal minimum moisture observed during summer drought and the maximum value measured during winter saturation \cite{Ott2025}. All parameters are treated as site- and depth-specific, enabling conversion of the dielectric response into an absolute PAW indicator.

\subsection{Multi-Channel Sensing and Crosstalk Compensation}
\label{sec:differentialSensing}
% ========= Full-width matrix equation =========
\begin{figure*}[!b]
\centering
\vspace{-18pt}
\begin{equation}
\label{eq:capMatrix}
\begin{bmatrix} 
    \underline{Y}_\text{p,a}(\varepsilon_\text{r},T) +\underline{Y}_\text{sb} +\underline{Y}_\text{off} 
    & -\underline{Y}_\text{p,a}(\varepsilon_\text{r},T) 
    & -\underline{Y}_\text{sb} \\[4pt]
    \underline{Y}_\text{p,a}(\varepsilon_\text{r},T) 
    & -\underline{Y}_\text{sa} -\underline{Y}_0(\varepsilon_\text{r}) -\underline{Y}_\text{p,a}(\varepsilon_\text{r},T) 
    & \underline{Y}_0(\varepsilon_\text{r}) \\[4pt]
    \underline{Y}_\text{sb} 
    & \underline{Y}_0(\varepsilon_\text{r}) 
    & -\underline{Y}_\text{sb}-\underline{Y}_0(\varepsilon_\text{r})-\underline{Y}_\text{p,b}(\varepsilon_\text{r},T)
\end{bmatrix}
\begin{bmatrix}
    \varphi_1 \\ \varphi_2 \\ \varphi_3
\end{bmatrix}
=
\begin{bmatrix}
    \underline{i}_\text{Q} \\ 0 \\ 0
\end{bmatrix}
\end{equation}
\end{figure*}
% Phi equation
\begin{figure*}[!b]
\centering
\vspace{-18pt}
\begin{equation}
\label{eq:phi1}
\varphi_1 = 
\frac{\underline{i}_\text{Q}}{\underline{a}(\varepsilon_\text{r},T)+\underline{Y}_\text{sb}+\underline{Y}_\text{off}}
\Biggl(
1 -
\frac{
\bigg(
\underline{Y}_\text{sb} -
\underline{Y}_\text{p,a}(\varepsilon_\text{r},T)
\frac{
\underline{a}(\varepsilon_\text{r},T)\underline{Y}_0(\varepsilon_\text{r}) +
\underline{Y}_\text{sb}\underline{Y}_\text{p,a}(\varepsilon_\text{r},T)
}{
\underline{a}(\varepsilon_\text{r},T)\underline{b}(\varepsilon_\text{r},T) +
\underline{Y}_\text{p,a}(\varepsilon_\text{r},T)^2
}
\bigg)^2
}{
\underline{a}(\varepsilon_\text{r},T)\underline{c}(\varepsilon_\text{r},T) +
\underline{Y}_\text{p,a}(\varepsilon_\text{r},T)^2 -
\frac{
\big(
\underline{a}(\varepsilon_\text{r},T)\underline{Y}_0(\varepsilon_\text{r}) +
\underline{Y}_\text{sb}\underline{Y}_\text{p,a}(\varepsilon_\text{r},T)
\big)^2
}{
\underline{a}(\varepsilon_\text{r},T)\underline{b}(\varepsilon_\text{r},T) +
\underline{Y}_\text{p,a}(\varepsilon_\text{r},T)^2
}
}
-
\frac{1}{
\frac{
\underline{a}(\varepsilon_\text{r},T)\underline{b}(\varepsilon_\text{r},T)
}{
\underline{Y}_\text{p,a}(\varepsilon_\text{r},T)^2
}
+1
}
\Biggr)
\end{equation}
\end{figure*}
% ========= Full-width auxiliary definitions =========
\begin{figure*}[!b]
\centering
\vspace{-18pt}
\begin{equation}
\label{eq:auxDefinitions}
\underline{a}(\varepsilon_\text{r},T)= \underline{Y}_\text{p,a}(\varepsilon_\text{r},T)+\underline{Y}_\text{sb}+\underline{Y}_\text{off},\quad
\underline{b}(\varepsilon_\text{r},T)= -\underline{Y}_\text{sa} -\underline{Y}_0(\varepsilon_\text{r})-\underline{Y}_\text{p,a}(\varepsilon_\text{r},T),\quad
\underline{c}(\varepsilon_\text{r},T)= -\underline{Y}_\text{sb} -\underline{Y}_0(\varepsilon_\text{r})-\underline{Y}_\text{p,b}(\varepsilon_\text{r},T)
\end{equation}
\end{figure*}
In multi-channel probes, parasitic coupling causes sensing-depth distortion and temperature drift \cite{Maier2023}, mitigated by alternating electronic isolation of the electrodes using two electronically controlled switches that separate parasitic from true soil capacitances.
The resulting behaviour is modelled using the equivalent circuit in Fig.~\ref{fig:ESBmodelling}.  
Nodal analysis with node potentials $\varphi_0$ to $\varphi_3$ (with $\varphi_0=0$ as ground) yields the Maxwell capacitance matrix in Eq.~\ref{eq:capMatrix}, with $\underline{Y}$ being the complex admittances of the respective capacitances \cite{Escriba2020}.
The measured capacitance follows from Ohm’s law via 
\( C_\mathrm{m}(\varepsilon_\text{r},T)=\underline{i}_\mathrm{Q}\big/ \big(j\omega(\varphi_1-\varphi_0)\big) \), 
where \( j \) is the imaginary unit.
Gaussian elimination yields the analytical solution in Eq.~\ref{eq:phi1}. 
The auxiliary variables are defined in Eq.~\ref{eq:auxDefinitions}. 
By evaluating $C_\text{m}(\varepsilon_\text{r},T)$ in all switch states (both closed, one open, both open) and computing the corresponding limit cases (capacities approaching 0 or $\infty$), the individual parasitic contributions can be isolated. 
Combining these configurations enables differential cancellation of all stray effects, yielding the corrected differential capacitance in Eq.~\ref{eq:CdiffResult}.
This differential activation approach isolates the effective soil capacitance $C_0(\varepsilon_\text{r})$ and eliminates temperature- and geometry-dependent parasitic influences, enabling accurate multi-depth measurements under field conditions.
\begin{multline}
\label{eq:CdiffResult}
C_\text{diff}(\varepsilon_\text{r}) =\;
\lim_{\substack{\underline{Y}_\text{sa}\to\infty\\\underline{Y}_\text{sb}\to\infty}} 
  C_\text{m}(\varepsilon_\text{r},T)
-\lim_{\substack{\underline{Y}_\text{sa}\to0\\\underline{Y}_\text{sb}\to\infty}} 
  C_\text{m}(\varepsilon_\text{r},T) \\[4pt]
-\lim_{\substack{\underline{Y}_\text{sa}\to\infty\\\underline{Y}_\text{sb}\to0}} 
  C_\text{m}(\varepsilon_\text{r},T) 
+\lim_{\substack{\underline{Y}_\text{sa}\to0\\\underline{Y}_\text{sb}\to0}} 
  C_\text{m}(\varepsilon_\text{r},T)
= C_0(\varepsilon_\text{r}) .
\end{multline}

\section{Results}
\subsection{Design and Construction Summary}
The differential scheme in Eq.~\ref{eq:CdiffResult} suppresses parasitic coupling and temperature drift by alternating electrode switch states. 
Low off-state capacitance and switch placement directly at the electrode feed are essential to limit stray effects \cite{Maier2023}. 
Optically coupled MOSFET bridges were selected for their galvanic isolation. 
The probe uses a four-layer FR4 PCB (shielded traces, gold-plated electrodes) in a polycarbonate housing, reducing compaction and improving robustness. 
Low-duty-cycle operation yields milliwatt-range average power consumption, enabling long-term battery-powered deployments.

\subsection{Volume Sensitivity}
\label{subsec:volumeSensitivity}
The dynamic range of the differential measurement $C_\text{diff}(\varepsilon_\text{r})$ is roughly five times smaller than that of the direct measurement $C_\text{m}(\varepsilon_\text{r},T)$ (Table~\ref{tab:dynRangeVolSel}), based on air and water measurements at room temperature for three depths. 
For sensing-volume characterization, a \SI{1.2}{\liter} water-filled latex balloon was placed concentrically in individual probe depths, while the surrounding volume was kept dry. 
Air, balloon-filled, and immersed states were sampled for \SI{5}{\minute} and averaged.
The relative volume contribution was computed as the ratio between the balloon-induced and full-scale signal changes. 
The differential setup captured \SI{93} - \SI{96}{\percent} of the dynamic range, compared to \SI{63} - \SI{68}{\percent} for direct measurements, confirming improved spatial confinement of the measurement.
\begin{table}[!t]
\centering
\caption{Dynamic Range and Volume Selectivity of Sensor Design}
\label{tab:dynRangeVolSel}
\vspace{-6pt}
\begin{tabular}{c||cc||cc}
\hline
\textbf{Channel} 
& \multicolumn{2}{c||}{\textbf{Dynamic Range}} 
& \multicolumn{2}{c}{\textbf{Volume Selectivity}} \\ 
\cline{2-5}
& $C_\mathrm{m}(\varepsilon_\text{r},T)$ 
& $C_\mathrm{diff}(\varepsilon_\text{r},T)$
& $C_\mathrm{m}(\varepsilon_\text{r},T)$
& $C_\mathrm{diff}(\varepsilon_\text{r},T)$ \\ \hline\hline
\SI{-30}{\centi\meter} & \SI{60.4}{\pico\farad} & \SI{13.8}{\pico\farad} & \SI{63}{\percent} & \SI{93}{\percent} \\
\SI{-60}{\centi\meter} & \SI{75.7}{\pico\farad} & \SI{16.2}{\pico\farad} & \SI{63}{\percent} & \SI{95}{\percent} \\
\SI{-90}{\centi\meter} & \SI{67.6}{\pico\farad} & \SI{14.5}{\pico\farad} & \SI{68}{\percent} & \SI{96}{\percent} \\ \hline
\end{tabular}
\vspace{-8pt}
\end{table}

\subsection{Temperature Sensitivity}
\label{subsec:tempSensitivity}
\begin{figure}[!t]
  \centering
  \includegraphics[width=0.42\textwidth]{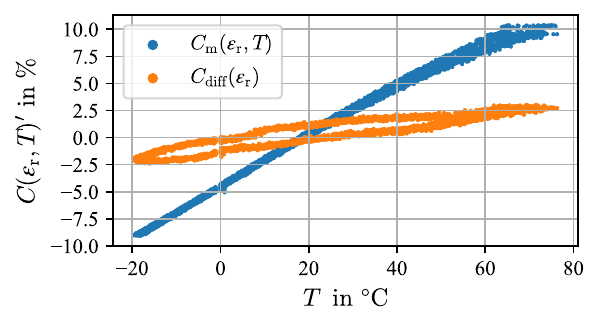}
  \vspace{-13pt}
  \caption{Temperature sensitivity test results. The normalized capacitance is shown for the \SI{-30}{\centi\meter} channel during multiple heating and cooling loops.}
  \label{fig:tempStab}
  \vspace{-12pt}
\end{figure}
The temperature dependence of the sensor response was evaluated using a programmable climate chamber. 
Multiple temperature loops from \SIrange{-20}{70}{\celsius} were conducted, each lasting three hours to ensure thermal equilibrium across components.
Fig.~\ref{fig:tempStab} shows the normalized capacitance for both direct and differential modes. 
The differential configuration exhibits approximately four-fold lower temperature sensitivity and minor hysteresis compared to the conventional measurement, demonstrating improved thermal stability.  

\subsection{Repeatability in Real Soil}
\label{subsec:repeatabilityInSoil}
\begin{figure}[!t]
  \centering
  \includegraphics[width=0.48\textwidth]{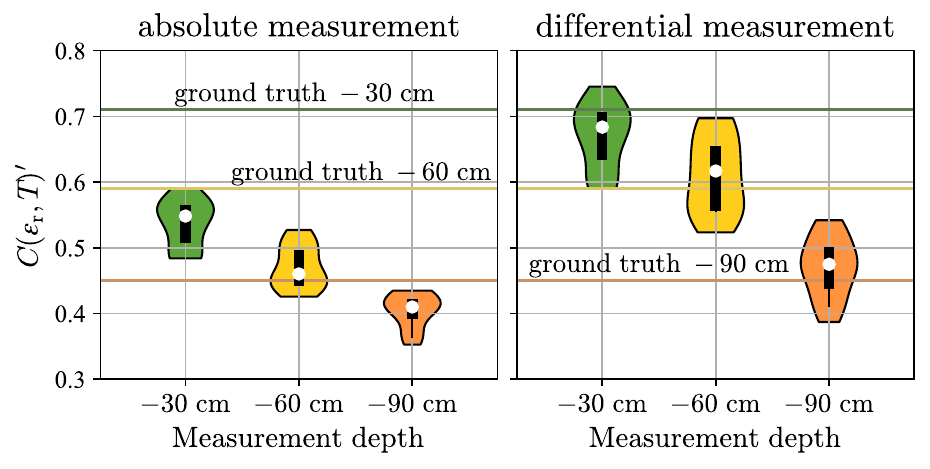}
  \vspace{-10pt}
  \caption{Repeatability test results in sandy soil showing normalized capacitance distributions for direct and differential measurements with reference values.}
  \label{fig:fieldTestViolin}
  \vspace{-10pt}
\end{figure}
The repeatability of the improved soil moisture sensor was evaluated in a sandy field during winter in central Europe. 
Insertion was standardized with a steel replica of the probe with a ground plate used to pre-form slits before placing the sensor.
It was inserted 15 times within \SI{4}{\meter^2}, and the resulting capacitance values were compared with gravimetric references from oven-dried samples.
Fig.~\ref{fig:fieldTestViolin} shows the normalized capacitances as violin plots. 
$C_\text{m}(\varepsilon_\text{r},T)$ showed \SI{5}{\percent} interquartile range (IQR) and up to \SI{15}{\percent} deviation, while $C_\text{diff}(\varepsilon_\text{r},T)$ showed \SI{8}{\percent} IQR and only \SI{3}{\percent} deviation from ground truth, indicating improved accuracy and robustness.

\subsection{Field Sensor Deployment and Performance}
\begin{figure}[!t]
    \centering
    \includegraphics[width=\linewidth]{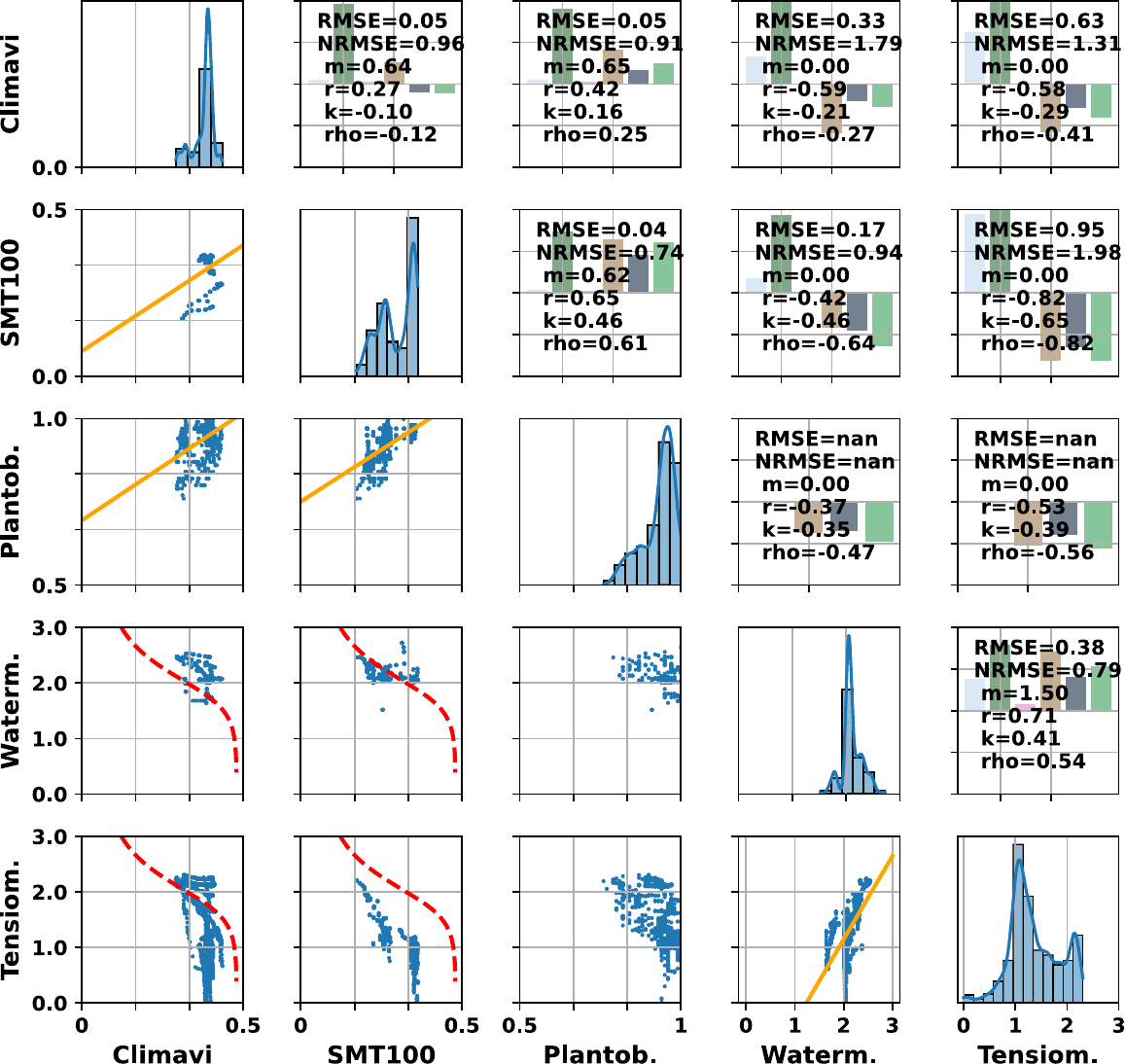}
    \vspace{-20pt}
    \caption{Pairwise correlations for tree 3 at \SI{-30}{\centi\meter}. Scatter plots show agreement between co-located sensors. Correlation \& error metrics shown for each pair. Linear fits in orange and van Genuchten curves in red.}
    \label{fig:correlationPlotLocation360}
    \vspace{-5pt}
\end{figure}
\begin{figure}[!t]
    \centering
    \includegraphics[width=0.45\textwidth]{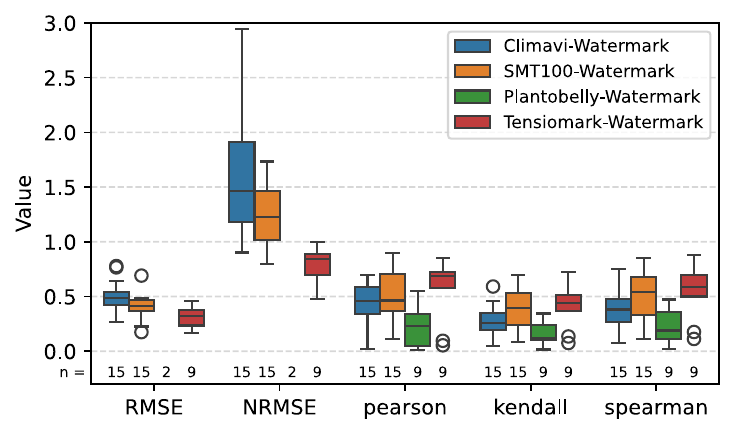}
    \vspace{-10pt}
    \caption{Aggregated performance metrics (RMSE, NRMSE, and correlation coefficients) for all watermark sensor pairs across all locations and depths. Sample size per pair (n) is indicated below each box.}
    \label{fig:group4MetricsBoxplot}
    \vspace{-15pt}
\end{figure}

A field study was conducted on five urban street trees in Hanover, Germany, instrumented with up to five sensor types. 
Climavi (proposed design), SMT100 \cite{Bogena2017}, Plantobelly \cite{Plantobelly2025}, Watermark \cite{Shock2002}, and Tensiomark \cite{Babaeian2019b} sensors were installed at three depths (root ball, 30 cm, 60 cm). 
The first three measured VWC via permittivity, while Watermark and Tensiomark measured soil water tension. 
Given the established role of suction-tension sensors as practical references \cite{Lo2020, Perea2013}, Watermark served as benchmark.  
Temporally aligned measurements (maximum offset 50 min) were analyzed for each co-located sensor pair. 
Linear regression was applied to matching domains (e.g., VWC–VWC or pF–pF), and RMSE and NRMSE (RMSE/$\sigma_\text{ref}$) were computed. 
For cross-principle comparisons, the Van Genuchten model \cite{vanGenuchten1980} served as the regression reference. 
Pearson, Kendall, and Spearman coefficients quantified correlations. 
Results are shown in Figs.~\ref{fig:correlationPlotLocation360} and \ref{fig:group4MetricsBoxplot} across all trees and depths, following \cite{Jackisch2020}. 
All data, analysis code and processing pipeline are openly available in \cite{Maier2025Repo}. 

\section{Discussion and Conclusion}
This work introduces a differential sensing approach for soil moisture measurement combined with a PAW computation method. 
Two optically coupled semiconductor switches alternately isolate the electrodes, thereby suppressing superposed parasitic capacitances previously reported in \cite{Maier2023}, as confirmed by the equivalent-circuit model and closed-form analysis.

\textit{Dynamic range.} Although differential operation reduces the electrical dynamic range by approximately a factor of five (Table~\ref{tab:dynRangeVolSel}), the remaining span remains above the converter noise floor, corresponding to $<\SI{1}{\percent}$ VWC resolution and $\approx\SI{5}{\percent}$ deviation in field tests (Fig.~\ref{fig:fieldTestViolin}), compared to up to \SI{15}{\percent} for direct measurements. 
Quantization and long-term drift effects remain below seasonal variability (Fig.~\ref{fig:tempStab}), so improved volume confinement and thermal stability outweigh the reduced span.
The improved volume confinement and thermal stability therefore outweigh the reduced dynamic range in practical deployments.

\textit{Volume sensitivity.} The water-balloon experiment confirms that conventional direct measurements include contributions from outside the intended sensing region, whereas the differential scheme confines the effective sampling volume to the plate-to-plate space. 
This addresses the issue that small or poorly defined sensing volumes cause spatial variability during installation \cite{Bell1987, Jackisch2020, Dean1987, Ranjan1997}.
The findings align with confined-volume and compartmental/submersion tests \cite{Limsuwat2009, Nissen2003, Robinson2000} and indicate effective crosstalk suppression.

\textit{Temperature stability.} The absolute measurement exhibits a thermal response consistent with FR4 stack-up and dielectric behaviour \cite{Azoteq2023}. 
Differential operation markedly lowers the apparent temperature gradient. 
Residual drift can be attributed to the properties of the insulation material (polycarbonate) \cite{Vaclav2001} and incomplete thermal equilibrium. 
In field deployments, temperature-dependent water permittivity will slightly add to this effect \cite{Wraith1999, Owen1961, Nieberding2023}. 
TDR systems typically report permille degree sensitivities \cite{Decagon2009, Pepin1995}. 
The differential scheme enhances thermal robustness in the system response.

\textit{Repeatability \& field evaluation.} In sandy soil, the differential mode reduces the systematic offset to gravimetric ground truth with only a modest precision trade-off, consistent with its reduced dynamic range (Fig.~\ref{fig:fieldTestViolin}). Under urban conditions, cross-sensor comparisons show competitive agreement. Furthermore, no significant performance degradation due to aging or moisture ingress was observed during the field deployments. These results align with laboratory improvements in volume confinement and thermal stability, and they compare favorably to the variability reported for existing systems. \cite{Jackisch2020, Walker2004}.

\textit{Conclusion.} The optical–semiconductor-based differential probe substantially mitigates multi-channel crosstalk and improves thermal stability. It also better confines the sampling volume, enhancing accuracy across depths. Combined lab and field evidence suggest a practical path toward more reliable irrigation decisions in agricultural and urban settings. Future work will target precision gains (materials/geometry, signal processing) and systematic evaluation under varying soil properties without sacrificing the demonstrated robustness.

\section*{Acknowledgment}
The authors thank all project partners involved in the soil characterization and field measurement campaigns.
\vspace{-3pt}

\bibliographystyle{IEEEtran}
\bibliography{sample-base.cleaned}

\end{document}